\documentclass[aps,showpacs,preprintnumbers,amsmath,amssymb, 
twocolumn, tightenlines,
]{revtex4}

\usepackage[dvips]{graphicx}\usepackage{bm}
\usepackage{amsmath}
\usepackage{dcolumn}
\usepackage{bm}

\begin{document}

\bibliographystyle{apsrev} 

\title {Charge density of a positively charged vector boson may be negative}

\author{V.V.Flambaum} \email[Email:]{flambaum@phys.unsw.edu.au}
\author{M.Yu.Kuchiev} \email[Email:]{kuchiev@phys.unsw.edu.au}

\affiliation{School of Physics, University of New South Wales, Sydney
  2052, Australia}
 \date{\today}

    \begin{abstract} 
The charge density of vector particles, for example $W^{\pm}$, may change sign. 
The effect manifests itself even for a free  propagation; when the energy of the $W$-boson 
satisfies $\varepsilon > \sqrt{2} m$ and the standing-wave is considered the charge density oscillates in space.
The   charge density of $W$ also changes sign in a close vicinity of a Coulomb center.
The dependence of this effect on the $g$-factor for an arbitrary vector boson, for example $\rho$-meson , is discussed. 
An origin of this surprising effect is traced to the electric quadrupole
 moment and spin-orbit interaction of vector particles.
Their contributions to the current have a polarization nature.
 The charge density of this current,
 $\rho_\mathrm{\,Pol}=-\nabla \cdot \mathrm{P}$, where 
 $\mathrm{P}$ is an effective polarization vector that depends on the
 quadrupole moment and spin-orbit interaction, oscillates in space,
 producing zero contribution to the total charge.
    \end{abstract}
    

    \maketitle


We show that the charge density of vector bosons, $W$ bosons in particular, can change sign. The effect manifests itself either for sufficiently high energy of the $W$-boson, or for sufficiently strong attractive potential. The origin of this effect is related to the electric quadrupole moment and spin-orbit
 interaction of vector mesons.   

The Proca theory \cite{proca_1936} presumes that the magnetic g-factor of vector bosons 
is $g=1$. The Corben-Schwinger \cite{corben-schwinger_1940} formalism allows one to describe vector bosons with an arbitrary value of the $g$-factor. In particular, it is applicable for the case of $W$ bosons in the Standard Model, which have $g=2$, see e. g. \cite{weinberg2-2001}. A close correspondence between the Corben-Schwinger formalism and the gauge theory was discussed in \cite{schwinger_1964}, its relation to the Standard Model was discussed in \cite{kuchiev-flambaum-06,flambaum-kuchiev-06}. 

Following \cite{corben-schwinger_1940} let us describe a vector boson propagating in an electromagnetic field by the Lagrangian
\begin{eqnarray}
      \nonumber
      {\mathcal L} &=& -\frac{1}{2} \,W_{\mu\nu} W^{\mu\nu}+ m^2 \,W_\mu^+ W^\mu +
\\ \label{W}
  &&+\,i\, e \,(g-1)  F^{\mu\nu} \,W^+_\mu W_\nu ~.
    \end{eqnarray}
Here  $m$ is the mass of a vector boson, $\nabla_\mu=\partial_\mu +i e A_\mu$, $F^{\mu\nu}=\partial^\mu A^\nu-\partial^\nu A^\mu$,
 $W_{\mu\nu}=\nabla_\mu W_\nu-\nabla_\nu W_\mu$
 and the magnetic g-factor
is arbitrary. The Lagrangian (\ref{W}) gives the following wave equations
\cite{corben-schwinger_1940} 
	\begin{eqnarray}
      \label{wave}
      \left( \nabla^2+m^2\right) W^\mu
      +  i e g \,F^{\mu\nu}\,W_\nu-  
      \nabla^\mu \nabla^\nu \,W_\nu =0~.
    \end{eqnarray}
    Taking a covariant derivative in Eq.(\ref{wave}) one finds also
\cite{corben-schwinger_1940}
    \begin{eqnarray}
      \label{Lgauge}
      m^2\nabla_\mu W^\mu+ie (g-1)\,J_\mu W^\mu
-ie\frac{g-2}{2}F^{\mu\nu}W_{\mu\nu}=0~,
    \end{eqnarray}
    where 
      $J^\mu=\partial_\nu F^{\nu\mu}$
    is the external current, which creates the external field
    $F^{\nu\mu}$. Eq.(\ref{Lgauge}) guarantees that among four components of the vector $W_\mu$ 
    only three are independent, precisely how it should be for a massive
 particle. From Eq.(\ref{W}) one also derives the current of vector bosons
 $j_\mu$ \cite{corben-schwinger_1940}
    \begin{eqnarray}
      \label{js}
      j_{\phantom{\,}\mu}=ie\Big(W^{+\nu} W_{\mu\nu} +(g-1)\partial^\nu (W^+_\mu W_\nu) -c.c.\Big)~.
 \end{eqnarray}
Here $c.c.$ refers to two complex conjugated terms.
Let us start from the simplest case of a free motion, when
Eq.(\ref{js}) gives the following charge density for vector particles
    \begin{eqnarray}
 \nonumber
     \rho  \equiv j_{\phantom{\,}0} = e\Big(2 \varepsilon (\boldsymbol{W}^+
     \! \cdot \!\boldsymbol{W}+(g-1)W^+_0 W_0)\\
      \label{rofree}
+\,ig(\boldsymbol{W}^{+}\!\cdot \!
     \boldsymbol{ \nabla} {W_0}-\boldsymbol{W}\!\cdot\!
     \boldsymbol{ \nabla} {W_0^+}) \Big)~.
    \end{eqnarray}
 Eq. (\ref{Lgauge}) gives
 $W_0=-i (\boldsymbol{ \nabla}\cdot\boldsymbol{W})/ \varepsilon$. For the plane
wave  $\boldsymbol{W}=\boldsymbol{C}\exp(i\boldsymbol{p}\cdot
     \boldsymbol{r})$ one observes a conventional, positively defined  charged density
\begin{eqnarray}
\label{plane}
\rho=(2e/ \varepsilon)\big(|\,C\,|^2\varepsilon^2-
|\,\boldsymbol{C}\cdot\boldsymbol{p}\,|^2\big)~,
\end{eqnarray}
 where $p=(\varepsilon^2-m^2)^{1/2}$ is the momentum.
 However, for the standing wave
$\boldsymbol{W}=\boldsymbol{C}\sin(\boldsymbol{p}\cdot\boldsymbol{r})$ an unusual effect takes place, the sign of the charge density is not necessarily fixed.
 The charge density in this case reads
\begin{eqnarray}
\nonumber
\rho&=& \frac{2e}{\varepsilon}\Big((g-1)|\,\boldsymbol{C}\cdot\boldsymbol{p}\,|^2\\
\label{stand}
  &-&\big( \,(2g-1)|\,\boldsymbol{C}\cdot\boldsymbol{p}\,|^2  -|\,C\,|^2\varepsilon^2\,\big)
\sin^2 (\boldsymbol{p}\cdot\boldsymbol{r})\Big).
\end{eqnarray}
For the longitudinal polarization,
 $|\,\boldsymbol{C}\cdot\boldsymbol{p}\,|=|\,C\,|p$, the density is
\begin{eqnarray}
\label{standl}
\rho\!= \!\frac{2e|C|^2\!}{\varepsilon}\Big((g\!-\!1)p^2\!
  -\!\big((2g-1)p^2 \! - \!\varepsilon^2\big)
\sin^2 (\boldsymbol{p}\!\cdot\!\boldsymbol{r})\Big).
\end{eqnarray}
We see that for $g>1$ and energy  $\varepsilon >m \sqrt{g/(g-1)}$ the charge density
 may take negative values. In the case of the $W$-boson,  $g=2$,
 the change of sign appears
for the energies $\varepsilon > \sqrt{2} m$.
 The minimum of density corresponds to
 $\sin^2 (\boldsymbol{p}\cdot\boldsymbol{r})=1$.
  In the Proca case $g=1$ the sign of the charge density is fixed.
However, for $g<1$ the charge density may be negative again.
 The minimum of the density in this case corresponds to
 $\sin (\boldsymbol{p}\cdot\boldsymbol{r})=0$. As we will show
below this behavior is explained by the contribution of the
electric quadrupole moment  of the vector particle, $Q \propto (g-1)e/m^2$.

 The motion of the $W$-boson (g=2) in the attractive Coulomb
 potential  $U=- Z\alpha/r$ gives another example. Take the most
 interesting case
of the total angular momentum $j=0$. Then 
${ \bf W }( {\bf r} ) = { \bf n }\,v(r)~$, where ${\bf n}={\bf r}/r~$ and
$v(r)$ is the radial wave function
\cite{corben-schwinger_1940,kuchiev-flambaum-06}.
Eq. (\ref{js}) gives the density of charge for this state,
\begin{equation}
\rho= 2 e\Big ((\varepsilon-U)(v^2+w^2)+2 v \frac{dw}{dr}\Big)~,	
	\label{rho1}
\end{equation}
where  $ w(r)=(v'(r)+2v(r)/r)/(\varepsilon-U)$
  \cite{corben-schwinger_1940,kuchiev-flambaum-06}.
At large distance we may neglect the potential, and the wave equation
has usual spherical-wave solution 
$v(r)\approx \sin (p \,r +\delta)/r$ or $\exp (-\kappa r)/r$. Again,
 the change of sign appears
for energies $\varepsilon > \sqrt{2} m$. The result looks even more
interesting for a charge density in a vicinity
of the Coulomb center. The most singular
 term in the Coulomb solution for W-boson is 
 $v(r) \sim  r^{\gamma-3/2}$ where
 $\gamma=(1/4- Z^2\alpha^2)^{1/2}$
 \cite{corben-schwinger_1940,kuchiev-flambaum-06} .
The charge density in this case is
$\rho \propto - e\, r^{2 \gamma -4}$
 \cite{corben-schwinger_1940,kuchiev-flambaum-06}.
Thus, the charge
  in a vicinity of the Coulomb center  has the ``wrong" sign.
Moreover,  the integral of the charge density
is divergent and this wrong-sign charge is 
 infinite. In \cite{kuchiev-flambaum-06} we showed
that the Coulomb problem for $W$-boson is saved by the fermion
vacuum polarization which produces the impenetrable potential barrier
near the origin, $U_{eff}\sim Z^2 \alpha^3/m^3 r^4$ \cite{vacuum}.
As a result, the $W$-boson charge density decreases exponentially and vanishes
at $r=0$,  which makes the  charge with wrong-sign near the origin finite.

To make a physical nature of this phenomenon more transparent
consider the nonrelativistic approximation for the Hamiltonian that
 describes a vector particle in a static electric field
(this Hamiltonian for $g=2$ was 
derived in \cite{kuchiev-flambaum-06})
         \begin{eqnarray}
         \label{hami}
      H_{ij}&=& \left( \frac{ \boldsymbol{ p}^2 }{2m}+U \right) 
      \delta_{ij}   
     - \frac{ \boldsymbol{ p}^4 }{8m^3}\,\delta_{ij}
     \\    \nonumber 
     &&-\frac{g-1}{2m^2}\,
     \big(  \,\boldsymbol{ F} \cdot (  \boldsymbol{ p \times S}_{ij} ) 
          -\nabla_i \nabla_j \,U \,\big)~,
    \end{eqnarray}
    Here $i,j=1,2,3$ label components of three-vectors, the nonrelativistic 
    Schredinger equation in this notation reads $H_{ij}\Phi_j=E \Phi_i$,
    $\boldsymbol{ S}$ is the spin, which operates on a vector
    $\boldsymbol{ V}$ according to $\boldsymbol{ S}_{ij}
    V_j=-i\epsilon_{ijk} V_k$, $U=eA_0$ and ${\bf F}=-\boldsymbol{\nabla} U$
 are the potential energy and the force produced by the field.
 The first and second terms in Eq.(\ref{hami}) describe the basic
 nonrelativistic approximation and 
    the relativistic correction to the mass respectively, the term with the spin
    gives  the spin-orbit interaction. 
    The last term in Eq.(\ref{hami}) includes two contributions:
the contact term $\propto \Delta  U \delta_{ij}/3$ and the quadrupole moment term
$\propto  \nabla_i \nabla_j U -\Delta U \,\delta_{ij}/3$. The corresponding density
 of the electric quadrupole moment equals 
 \begin{equation}
Q_{ij}=(g-1)\, \frac{e}{ m^2} \,\big (\,3
      \boldsymbol{ \Phi }_i^*
      \boldsymbol{ \Phi }_j-\delta_{ij}
      |  \boldsymbol{ \Phi } |^2\big)~,	
	\label{Q}
\end{equation}
  where
\begin{equation}
\label{wf}
\boldsymbol{\Phi}     \simeq \boldsymbol{W}+\boldsymbol{\nabla}\,
       ( \boldsymbol{\nabla} \cdot \boldsymbol{W} )/2m^2~,
\end{equation}
is the non-relativistic wave function introduced in \cite{kuchiev-flambaum-06}.

 Calculating a variation of 
    the matrix element $\langle\Phi|H|\Phi\rangle$ 
    of the Hamiltonian Eq.(\ref{hami}) with respect to the
 potential we find 
    the charge density $\rho _\mathrm{nr}$ of a vector particle
 in the nonrelativistic 
    approximation: 
\begin{eqnarray}
	\label{rho}
	\rho _\mathrm {nr}&=&
	\rho _\mathrm{C}+\rho _\mathrm{ S}+\rho _\mathrm{ Q}~,
	\\ \label{e} 
	\rho _{\rm C}
    &=& e \,\boldsymbol{\Phi}_i^*\boldsymbol{\Phi}_i~,
  \\ \label{S)} 
  \rho _{\rm S}&=&  
 (g-1)\,\frac{e}{2m^2}\,
    (\, \nabla_i \boldsymbol{\Phi}_i^*\nabla_j\boldsymbol{\Phi}_j-\nabla_j
     \boldsymbol{\Phi}_i^*\nabla_i\boldsymbol{\Phi}_j\,),~~~
     \\ \label{roQ}
     \rho _{\rm Q}
     &=&( g-1)\,\frac{e}{2m^2}\,\nabla_i\nabla_j\,(\boldsymbol{\Phi}_i^*\boldsymbol{\Phi}_j)~.
\end{eqnarray}
The term $\rho _{\rm C}$ here can be interpreted as a
 conventional nonrelativistic distribution of charge. The next term
 $\rho _{\rm S}$ originates from the spin-orbit interaction in
 Eq.(\ref{hami}). The term $\rho _{\rm Q}$ comes from the
 last  term in the Hamiltonian (\ref{hami}), which describes the 
 quadrupole and contact interactions of the boson.
The spin-orbit and quadrupole terms
 give zero contribution to the total charge because each one of them can
 be written as a divergence. For the quadrupole term $\rho^W_{\rm Q}$ this is
 evident from its definition.  The spin-orbit term  can be rewritten
 in such a way,
$\rho^W_{\rm S}\propto
 \nabla_i ( \boldsymbol{\Phi}_i^*\nabla_j\boldsymbol{\Phi}_j-
     \boldsymbol{\Phi}_j^*\nabla_j\boldsymbol{\Phi}_i)\Big)$, as to make it
 clear that it is a divergence as well.
Thus, the sum $\rho _{\rm S}+\rho _{\rm Q}$ may be presented as a divergence
 of a vector,
\begin{eqnarray}
		\label{div}
		&&\rho _{\rm S}+\rho _{\rm Q}\equiv 
		\,\rho _\mathrm{\,Pol}~
 = -\boldsymbol{\nabla} \cdot \boldsymbol{P} \,.
\end{eqnarray}
Therefore, it
can be looked at as a charge density  $\rho _\mathrm{\,Pol}$
 induced by an effective polarization which is described by the polarization
 vector $\boldsymbol{P}$ \cite{dipole}.
This vector can be identically rewritten
 in the following simple form 
\begin{eqnarray}
		\label{P}
		 &&\boldsymbol{P}=-(g-1) \,\frac{e}{m^2}\,\mathrm{Re}\, \big(\,\boldsymbol{\Phi}^*(\boldsymbol{\nabla}\cdot
		 \boldsymbol{\Phi})\,\big)
\end{eqnarray}
 Eq.(\ref{div}) explicitly shows that $\rho _{\rm \,Pol}$ does not contribute
 to the total charge of the vector boson, $ \int \rho _\mathrm{\,Pol} d^3r=0$. Consequently, $\rho _{\rm \,Pol}$ inevitably oscillates, changing sign.
Same conclusion holds  separately for $\rho _{\rm S}$ and $\rho _{\rm Q}$.

 In the low-energy region $\rho _{\rm \,Pol}$ is smaller than the conventional
 charge density $\rho _\mathrm{C}$, which has a definite sign. However, with
 increase of momentum $|\rho _{\rm \,Pol}/\rho _\mathrm{C}|$ is growing as
 $p^2/m^2$ or $1/r^2 m^2$  
since it depends on derivatives of the wave function. For sufficiently high
 energy 
(or large attractive potential) $\rho _{\rm \,Pol}$ may prevail, forcing the
 total charge density $\rho $ to change its sign as well. However, to make 
this argument rock solid for high energies, one has to consider a fully 
relativistic treatment, which is presented below, see Eqs.(\ref{ro})-(\ref{PP}).

To clarify the meaning of Eqs.(\ref{rho})-(\ref{P})
let us consider  specific cases for a low-energy vector particle.
 For the plane
wave  $\boldsymbol{W}=\boldsymbol{C}\exp(i\boldsymbol{p}\cdot
     \boldsymbol{r})$ the spin-orbit and quadrupole contributions
to the charge density vanish, $\rho _{\rm Q}=\rho _{\rm S}=0$, and the density
$\rho =\rho _{\rm C}=const$, in agreement with Eq. (\ref{plane}).
For the standing wave
$\boldsymbol{W}=\boldsymbol{C}\sin(\boldsymbol{p}\cdot\boldsymbol{r})$
we obtain the  positive non-relativistic charge density, 
zero spin density and oscillating quadrupole density
\begin{eqnarray}
	\label{stanC}
\rho _{\rm C}&=&e(|C|^2-|\boldsymbol{C}\cdot\boldsymbol{p}/m|^2)
\sin^2 (\boldsymbol{p}\cdot\boldsymbol{r})~,
\\ \label{stanS}
\rho _{\rm S}&=&0~,
\\
\label{stanQ}
\rho _{\rm Q}&=&e(g-1)|\boldsymbol{C}\cdot\boldsymbol{p}/m|^2
\cos(2\boldsymbol{p}\cdot\boldsymbol{r})~.
\end{eqnarray}
As expected, the quadrupole density
 increases with $p^2$. The total density agrees with
Eq.(\ref{stand}) for $\varepsilon \approx m$.

Finally, consider
a vector particle in the attractive Coulomb field in the state $2p_0$ ($j=0,~l=1$).
In this case
\begin{eqnarray}
\label{rhPhi}
\boldsymbol{\Phi}&=&C \boldsymbol{r }\exp(-kr/2)~,
\\ \label{rhC}
\rho _{\rm C}&=& eC^2 r^2\exp(-kr)~,
\\ \label{rhS}
\rho _{\rm S}&=&(g-1)eC^2(3\!-\!kr)\exp(-kr)/m^2~,
\\\label{rhQ}
\rho _{\rm Q}&=&(g-1)eC^2(6\!-\!4kr\!+\!k^2r^2/2)\exp(-kr)/m^2\!.\quad
\end{eqnarray} 
Here $k=Z\alpha m$, while $C$ describes the normalization of the wave function.
We see that the spin density and the quadrupole density are enhanced
by the factor $\propto 1/m^2 r^2$ at small distances in comparison with
 the non-relativistic charge density $\rho _{\rm C}\propto r^2$. 
Also, $\rho _{\rm S}$ and $\rho _{\rm Q}$ have the radial oscillations
and vanish after the radial integration,
 $\int \rho_Q \,r^2dr=\int \rho_S \,r^2dr=0$.

   Let us discuss now the relativistic case.
For simplicity consider propagation of the $W$-boson ($g=2$)
 in a static external field
 $U=eA_0,~{\bf A} =0$. In the region, where the external current is absent,
 $J_\mu=0$,  Eq.(\ref{Lgauge}) gives 
\begin{equation}
\label{w}
w\,\equiv \,i\, W_0=\boldsymbol{\nabla} \cdot \boldsymbol{W}/ (\varepsilon-U)~,	
\end{equation}
 and the charge density extracted from Eq.(\ref{js}) may be presented in the following form:
    \begin{eqnarray}
      \label{ro}
     \rho \,\,&= &\rho_{\rm P}+\rho_{CS}~,
     \\ \label{roI}
     \rho_{\rm P}\,&=& 2e\,\Big(  (\varepsilon-U)\,\boldsymbol{W}^+
     \!\!\cdot\!\boldsymbol{W}+ {\rm Re}\,(\boldsymbol{W}^{+}\!\!\cdot\!
     \boldsymbol{ \nabla} {w} \Big)~,
    \\ \label{roII}
    \rho_{\rm CS}&=& -\boldsymbol{ \nabla}\cdot \boldsymbol{P}_\mathrm{CS}~,
   \\
    \label{PP}
    \boldsymbol{P}_\mathrm{CS}&=& -2 e \,{\rm Re}\,(\boldsymbol{W}^{+} {w})~.
    \end{eqnarray}
Here $\rho_{\rm P}$  is related to the first 
two terms in the Lagrangian Eq.(\ref{W}) which were introduced by Proca
 \cite{proca_1936}. 
The density $\rho_{\rm CS}$ originates from the last term in Eq.(\ref{W}),
 which was firstly introduced by Corben and Schwninger \cite{corben-schwinger_1940} and is also present in the Standard Model (with $g=2$) \cite{kuchiev-flambaum-06}. 

The term $\rho_\mathrm{P}$ has a conventional, definite sign in the nonrelativistic approximation, when $\rho_\mathrm{P} \simeq \rho_\mathrm{C}/(2m)$, where $\rho_\mathrm{C}$ is defined in Eq.(\ref{e}). The factor $2m$ here accounts for the fact that we use a conventional notation, in which the normalization conditions for the relativistic and nonrelativistic wave functions differ by precisely this factor.
This term also keeps a conventional sign in the region of space where
the potential is weak, $ |U\ll \varepsilon $. The higher the energy, the wider this region is. In the ultrarelativistic limit $\varepsilon\rightarrow \infty$ the term $\rho_\mathrm{P}$ keeps definite sign almost everywhere, except close vicinities of any Coulomb center. 
However, since it is known that at in the vicinity of the Coulomb center the
the $W$-boson charge density decreases \cite{kuchiev-flambaum-06}, 
the small regions near the Coulomb centers give small contribution to the total charge.
Thus, for a wide range of energies and vast areas of space the term $\rho_\mathrm{P}$ in the charge density reveals conventional behavior. 

In contrast, the charge density $\rho_\mathrm{CS}$ in Eq.(\ref{roII}) is a divergence. As a result, it gives zero contribution to the total charge $\int \rho_\mathrm{CS}\,d^3r=0$, and inevitably oscillates in space, showing the variation of sign. In the nonrelativistic approximation it is reduced to the polarization charge density defined in Eq.(\ref{div}), $\rho_\mathrm{CS}\simeq \rho_\mathrm{Pol}/(2m)$.  
The higher the energy  the bigger are the spacial derivatives in Eqs.(\ref{roII}),(\ref{PP}), forcing $\rho_\mathrm{CS}$ to grow with increase of energy. In contrast, $\rho_\mathrm{P}$ cannot show similar growth due to a restriction related to 
the total charge of the $W$ boson, which equals the integral  $e=\int \rho_\mathrm{P}\,d^3r$, making large values of $\rho_\mathrm{P}$ impossible. We conclude that for high energies the term $\rho_\mathrm{CS}$ prevails in the charge density. It is inevitable therefore that the total charge  density of $W$ bosons changes its sing, revealing a ``wrong" sign in some areas of space. 

In summary, we verified that the charge density of vector particles can have
wrong sign; for example it can be negative for the $W^+$ boson.
The effect is related to the electric quadrupole moment and
spin-orbit interaction of the
vector boson which are proportional to $(g-1)$ and  originate from the
last term in the Lagrangian  Eq.(\ref{W}),
which was firstly introduced by Corben and Schwninger \cite{corben-schwinger_1940} and is also present in the Standard Model (with $g=2$) \cite{kuchiev-flambaum-06}. 

      
        This work was supported by the Australian Research Council.


\begin{thebibliography}{99}
\bibitem{vacuum}
 This barrier exists for S=1 and does not
exist for spin S=0 and S=1/2 particles ( where the effect of the vacuum
 polarization
becomes significant only at exponentially small distance
 $r \sim \exp(-1/\alpha)m^{-1}$). Note that this phenomenon also
exists for the repulsive Colomb  potential $U=Z\alpha/r$.
Indeed, the most singular terms in the wave equation (\ref{wave})
produce an effective potential
 $\sim -Z^2 \alpha^2/mr^2$ \cite{kuchiev-flambaum-06}.
 It creates the attraction and divergencies
for any sign of $Z$.
The  fermion vacuum polarization  produces the repulsive barrier
 $\sim Z^2 \alpha^3/m^3 r^4$ for any sign of $Z$ too.
This phenomenon for the vector bosons in non-Abelian theories $SU(2)$ and the
 Standard model $U(1) \times SU(2)$ is considered
 in \cite{flambaum-kuchiev-06}.
\end{thebibliography}
\end{document}